\documentclass[12pt]{article}
\usepackage{amssymb}
\usepackage{psfig}

\newcommand{\al}{\alpha}
\newcommand{\be}{\beta}
\newcommand{\pa}{\partial}

\newcommand{\si}{\sigma}

\newcommand{\tha}{\theta}

\newcommand{\rar}{\rightarrow}

\begin{document}
\begin{titlepage}

\begin{flushright}
M\'exico ICN-UNAM 11/00\\ LPT-ORSAY 00-79\\
October, 2000
\end{flushright}

\vskip 1.6cm

\begin{center}

{\LARGE Ion $H_2^+$ can dissociate in a strong magnetic field}

\vskip 0.8cm

Alexander V. Turbiner  \cite{byline} \cite{byline2},\\
 Laboratoire de Physique Theorique, Universite Paris-Sud, Orsay
 F-91405, France and
 Instituto de Ciencias Nucleares, UNAM, M\'exico\\[10pt]

 Juan Carlos L\'opez V. \cite{byline3},\\
Instituto de Ciencias Nucleares, UNAM, Apartado Postal 70-543,
      04510 M\'exico D.F., M\'exico\\[10pt]

   and\\[10pt]

 Antonio~Flores-Riveros \cite{byline4}\\
 Instituto de F\'{\i}sica, Benem\'erita Universidad Aut\'onoma de Puebla \\
Apartado Postal J-48, Puebla Pue. 72570, M\'exico

\vskip 0.8cm

\end{center}

\vskip 1.5cm

\centerline{Abstract}

\begin{quote}
  \hskip .5cm In framework of a variational method the molecular ion
  $H_2^+$ in a magnetic field is studied.  An optimal form of the
  vector potential corresponding to a given magnetic field (gauge
  fixing) is chosen variationally.  It is shown that for any magnetic
  field strength as well as for any orientation of the molecular axis
  the system $(ppe)$ possesses a minimum in the potential energy. The
  stable configuration always corresponds to elongation along the
  magnetic line.  However, for magnetic fields $B \gtrsim 5\times 10^{11}\,G$
  and some orientations the ion  $H_2^+$ becomes unstable decaying to
  $H\mbox{-atom} + p$.

\end{quote}



\end{titlepage}

Ion $H_2^+$ is the simplest one-electron molecular system, which is
more stable than hydrogen atom. It appears to be one of the most
studied problems in non-relativistic quantum mechanics. In particular,
the wealth of physical phenomena displayed by this system when placed
into a magnetic field becomes of a great importance in astrophysics,
solid state and plasma physics.  For instance, as the magnetic field
grows the system becomes more and more strongly bound and compact.
Such a behavior leads naturally to a guess that in spite of the huge
temperatures on neutron star surfaces their atmosphere can still
contain molecular objects \cite{Kadomtsev:1971,Ruderman:1971}. On the
other hand, a shrinking of the equilibrium distance between protons
with the growth of magnetic field, increases drastically the
probability of nuclear fusion \cite{Kher}. It is quite surprising that
such a shrinking is also accompanied by a change from ionic to
covalent character at $\sim 5\cdot 10^{11} G$ \cite{Larsen} (see also
\cite{Lopez:1997}). The goal of this Letter is two-fold. Firstly, to
show that the system $(ppe)$ has always a minimum and correspondingly
the molecular ion $H_2^+$ can exist for magnetic fields $\lesssim
4.414\cdot 10^{13} G$. Secondly, to demonstrate that for $B
\gtrsim 5\times 10^{11}\,G$ and for some orientations of the molecular
axis the ion becomes unstable dissociating to $H\mbox{-atom} + p$. The
variational method is used to study this problem.

The Hamiltonian which describes the $H_2^+$ molecular ion placed
in a uniform constant magnetic field directed along the $z-$axis,
${\cal B}=(0,0,B)$ (see, for example,  \cite{LL}) is given by
\begin{equation}
\label{Ham}
 {\cal H} = {\hat p}^2 + \frac{2}{ R} -\frac{2}{r_1} -\frac{2}{r_2}  +
({\hat p} {\cal A}) +  {\cal A}^2 \ ,
\end{equation}
(see Fig.1), where ${\hat p}=-i \nabla$ is the momentum. ${\cal A}$ is
a vector potential, which corresponds to the magnetic field $\cal
B$.

The vector potential for given magnetic field is defined
ambiguously, up to a gauge factor. Thus, the Hamiltonian is
gauge-dependent but not the observables. Since we are going to use
an approximate method for solving (\ref{Ham}) our energies can be
gauge-dependent (only the exact ones would be gauge-independent).
Hence one can choose the form of the vector potential in an
optimal way. Let us consider a certain one-parameter family of
vector potentials corresponding to the constant magnetic field $B$
(see for example \cite{Larsen})
\begin{equation}
\label{Vec}
  {\cal A}= B(-(1-\xi)y, \xi x, 0)\ ,
\end{equation}
where $\xi$ is the parameter to be fixed in a certain
optimal way. If $\xi=1/2$ we get a gauge called symmetric or
circular, while $\xi=0$ corresponds to an asymmetric gauge (see
\cite{LL}). By substituting (\ref{Vec}) into (\ref{Ham}) we arrive at
the Hamiltonian
\begin{equation}
\label{Ham.fin}
 {\cal H} = -{\nabla}^2 + \frac{2}{ R} -\frac{2}{r_1} -\frac{2}{r_2}  +
 i B[-(1-\xi)y\pa_x + \xi x\pa_y] +  B^2 [ \xi^2 x^2+ (1-\xi)^2 y^2] \ .
\end{equation}

The idea of choosing an optimal gauge has widely been exploited in
quantum field theoretical considerations. It has also been
discussed in connection with the problem at hand (see for instance
\cite{Schmelcher} and references therein). Perhaps, the first
constructive (and remarkable) attempt to apply this idea was made
by Larsen \cite{Larsen}. In his study of the ground state it was
explicitly shown that gauge dependence of energy can be quite
significant and even an oversimplified optimization procedure
improves the numerical results.

Our aim is to study the ground state of (\ref{Ham.fin}). It is not
difficult to see that there exists a certain gauge for which
Hamiltonian (\ref{Ham.fin}) has a real ground state eigenfunction.
This gauge will be here sought after and correspondingly deal with
{\it real} trial functions in our variational calculations. In
this case one can prove that the expectation value of the term
$\sim B$ in (\ref{Ham.fin}) vanishes when it is taken over any real
normalizable function. So, without loss of generality we can omit
this term in the Hamiltonian.  Finally, the recipe of our
variational study can be formulated as follows:
 \emph{Construct an adequate variational real trial function
  \cite{Tur}, which reproduces the original potential near Coulomb
  singularities and at large distances, where $\xi$ should be included
  as a parameter. Perform a minimization of the energy functional by
  treating the trial function's free parameters and $\xi$ on the same
  footing.}
In particular, such an approach enables one  to eventually find the
\emph{optimal} form of the Hamiltonian. The above recipe was
successfully applied in a previous study of $H_2^+$ in magnetic
field \cite{Lopez:1997} and led to predict the existence of the
exotic ion $H_3^{++}$ at $B \gtrsim 10^{11}\,G$
\cite{Turbiner:1999}.

One of the simplest trial functions satisfying the above-mentioned
criterion is
 \begin{equation}
 \label{tr:1}
 \Psi_1= {e}^{-\al_1  (r_1+r_2)
  - B  [\be_{1x} \xi x^2 + \be_{1y}(1-\xi) y^2] }\ ,
 \end{equation}
(cf. \cite{Lopez:1997}) where $\al_1,\be_{1x,1y}$ are variational
parameters. We here assume that $\xi \in [0,1]$ which is a
restriction that will later be justified. Actually, this is a
Heitler-London function multiplied by the lowest Landau orbital
associated with the gauge (\ref{Vec}). Presumably this function
describes internuclear distances near the equilibrium and a
covalent character. Another possible trial function is
 \begin{equation}
 \label{tr:2}
 \Psi_2= \bigg({e}^{-\al_2 r_1}+ {e}^{-\al_2 r_2}\bigg)
 {e}^{ -  B  [\be_{2x}\xi x^2 +\be_{2y}(1-\xi) y^2] }\ ,
 \end{equation}
(cf. \cite{Lopez:1997}) where $\al_2,\be_{2x,2y}$ are variational
parameters. This is a Hund-Mulliken function multiplied by the
lowest Landau orbital. One can assume that for a sufficiently
large internuclear distance $R$ this function dominates, thus
giving an essential contribution in this regime. Hence, it
describes an interaction of a hydrogen atom and a proton (charged
center), and can also describe a possible decay mode of $H_2^+$
onto them. There are two natural ways$-$linear and non-linear$-$to
incorporate the behavior of the system both near equilibrium and
at long distances in a single trial function.  The non-linear
interpolation is of the form
 \begin{equation}
 \label{tr:3-1}
 \Psi_{3-1}= \bigg({e}^{-\al_3 r_1-\al_4 r_2}+
 {e}^{-\al_3 r_2-\al_4 r_1}\bigg)
 {e}^{ - B  [\be_{3x}\xi x^2 +\be_{3y}(1-\xi) y^2] }\ ,
 \end{equation}
(cf. \cite{Lopez:1997}) where $\al_{3,4},\be_{3x,3y}$ are
variational parameters. This is a Guillemin-Zener function
multiplied by the lowest Landau orbital. The linear superposition
is given by
 \begin{equation}
 \label{tr:3-2}
 \Psi_{3-2}= A_1 \Psi_{1} + A_2 \Psi_{2}\ ,
 \end{equation}
where one of the parameters $A_{1,2}$ is kept fixed.
The final form of the trial function is a linear superposition of
functions (\ref{tr:3-1}) and (\ref{tr:3-2})
 \begin{equation}
 \label{trial}
 \Psi_{trial}= A_1 \Psi_{1} + A_2 \Psi_{2}+A_{3-1} \Psi_{3-1}\ ,
 \end{equation}
 where only two out of the three parameters $A$'s are variationally
 treated. The total number of variational parameters in (\ref{trial})
 is fourteen when $\xi$ is included.

It is easy to prove a general statement that if a system possesses
axial rotational symmetry (in our case it appears if the molecular
axis coincides with the magnetic line, $\tha=0^o$, see Fig. 1),
the optimal gauge corresponds to $\xi=1/2$ (symmetric or circular
gauge). It is precisely the gauge which was used in most of
previously performed $H_2^+$-studies. However, this is not the
case if $\tha \neq 0^o$. As an example one can see in Fig. 2 the
behavior of $\xi$ as a function of $\tha$ at $B = 10^{12} G$. It
is typical behavior for all studied magnetic fields both weak and
strong, up to the non-relativistic limit, $B=4.414 \times 10^{13}
G$. It justifies our above assumption with regard to the domain of
gauge parameter, $\xi \in [0,1]$.

We carried out extensive studies of the ground state $1\si_g$ of
$H_2^+$ for magnetic fields $B=0 - 4.414 \times 10^{13} G$ and
orientations ranging from $0^o$ (parallel configuration) to $90^o$
(perpendicular configuration). They turned out to be more accurate
than any available results yet obtained except for a domain of
small magnetic fields for the perpendicular configuration, where
the results by Wille \cite{Wille:1988} appear to be slightly
better
 \footnote{These results were obtained using a basis set expansion
including up to approximately 500 terms depending on the value of
magnetic field.}.
The detailed numerical analysis and comparison with available
calculations will be published elsewhere.

As previously obtained by other authors
\cite{Kher,Wille:1988,Larsen} we confirm quantitatively a natural
expectation that the parallel configuration is the most stable for
all magnetic fields $B \lesssim 4.414 \times 10^{13} G$, where
non-relativisitic considerations are valid. The total energy of
the molecular ion $H_2^+$ as a function of angle $\tha$ for
different magnetic fields, is shown in Fig. 3. For any magnetic
field in the region $B=0 - 4.414 \times 10^{13} G$ and any
orientation a well-pronounced minimum in the total energy is
attained at finite internuclear distances. This is in contradiction
with a statement by Khersonsky \cite{Kher} about the non-existence
of a minimum for some values of the magnetic field and
orientation. Perhaps, it it worth to emphasize that in that
article the variational study was carried out using  a trial
function which is almost coincident with that of eq.(\ref{tr:2}).
We can thus presume that the above statement is an artifact
arising from insufficient accuracy of calculations. The horizontal
line in Fig. 3 presents the hydrogen atom total energy in the
magnetic field (see \cite{Salpeter:1992}).  For magnetic fields
$B>1.8 \times 10^{11}~G$ the total energy of the atom becomes
lower than that of $H_2^+$ for angles larger than some critical
angle, $\tha_{cr}$. It points to the possible dissociation channel
$H_2^+ \rar H\mbox{-atom} + p$. The dependence of the critical
angle $\tha_{cr}$ on the magnetic field is shown in Fig. 4. It is
quite striking that dissociation occurs for a wider and wider
range of orientations as the magnetic field grows reaching $25^o
\lesssim \tha \leqslant 90^o$ for $B=4.414 \times 10^{13} G$.

\vskip 2. cm
\noindent {\Large\bf Acknowledgments}

This article is dedicated to the memory of B.B.~Kadomtsev.

A.T. wishes to thank K.G.~Boreskov (ITEP, Moscow) and B.I.~Ivlev
(IF-UASLP, Mexico) for useful discussions. A.T. wants to express a
gratitude to LPTHE (Orsay) for hospitality extended to him where this
work was finished.

This work was supported in part by DGAPA Grant \# IN120199
(M\'exico).

\newpage

\def\href#1#2{#2}

\begingroup\raggedright

\endgroup
\pagebreak


\begin{figure}[htbp]
\begin{center}
     \parbox{5in}{
     \psfig{figure=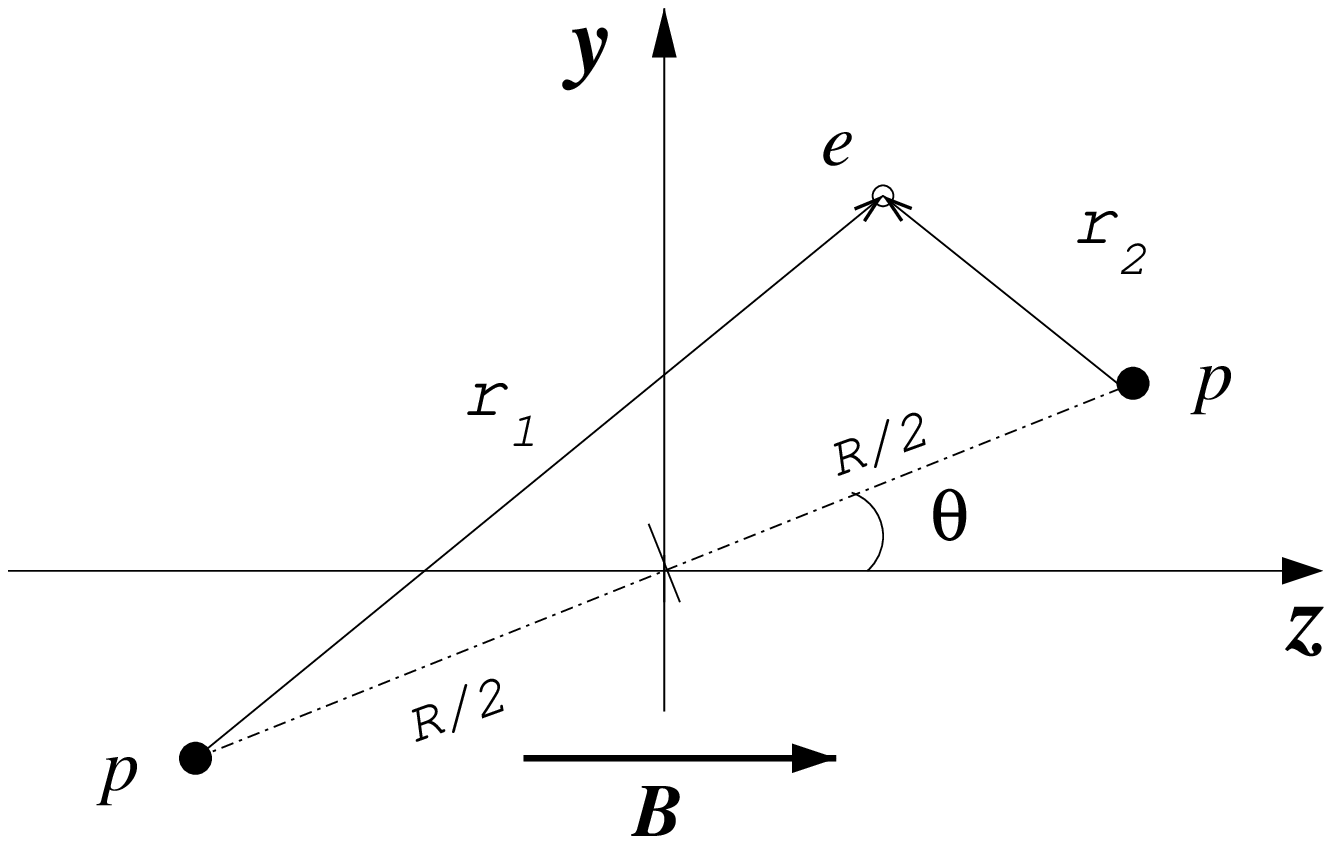,width=5.0in,angle=-90}}
     \caption{Geometrical setting  for the $H_2^+$ ion in a
      magnetic field directed along the $z$-axis.}
    \label{fig:1}
\end{center}
\end{figure}

\begin{figure}[htbp]
\begin{center}
     \parbox{5in}{
     \psfig{figure=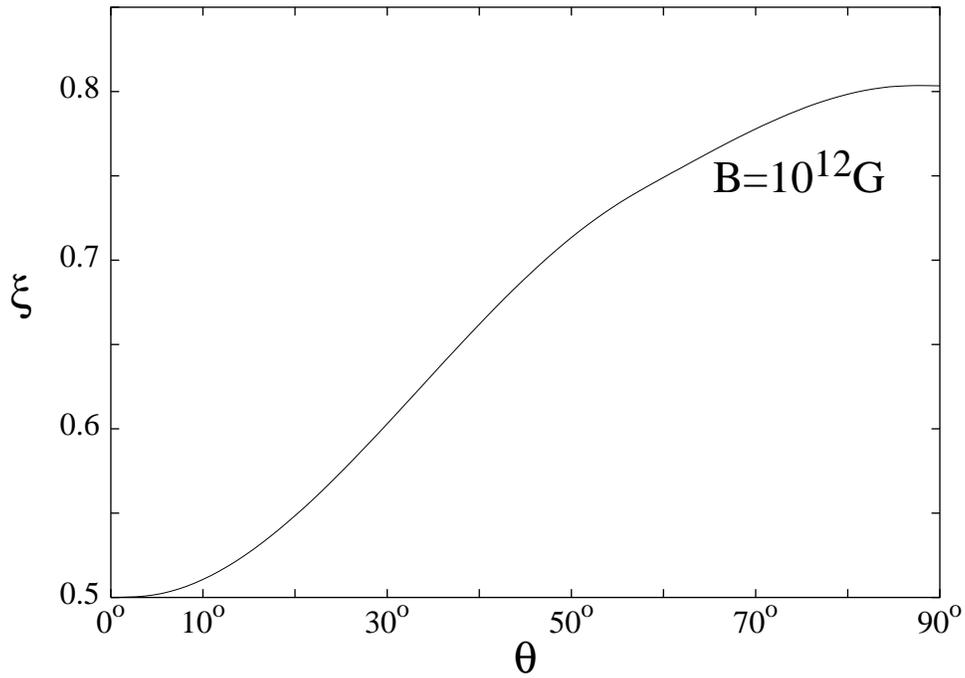,width=5.0in,angle=-90}}
     \caption{Gauge parameter as a function of the inclination angle
     for  $H_2^+$. The magnetic field $B=10^{12}$G was taken as an
     example. This dependence is found in the present study.}
    \label{fig:2}
\end{center}
\end{figure}

\begin{figure}[htbp]
\begin{center}
{\psfig{figure=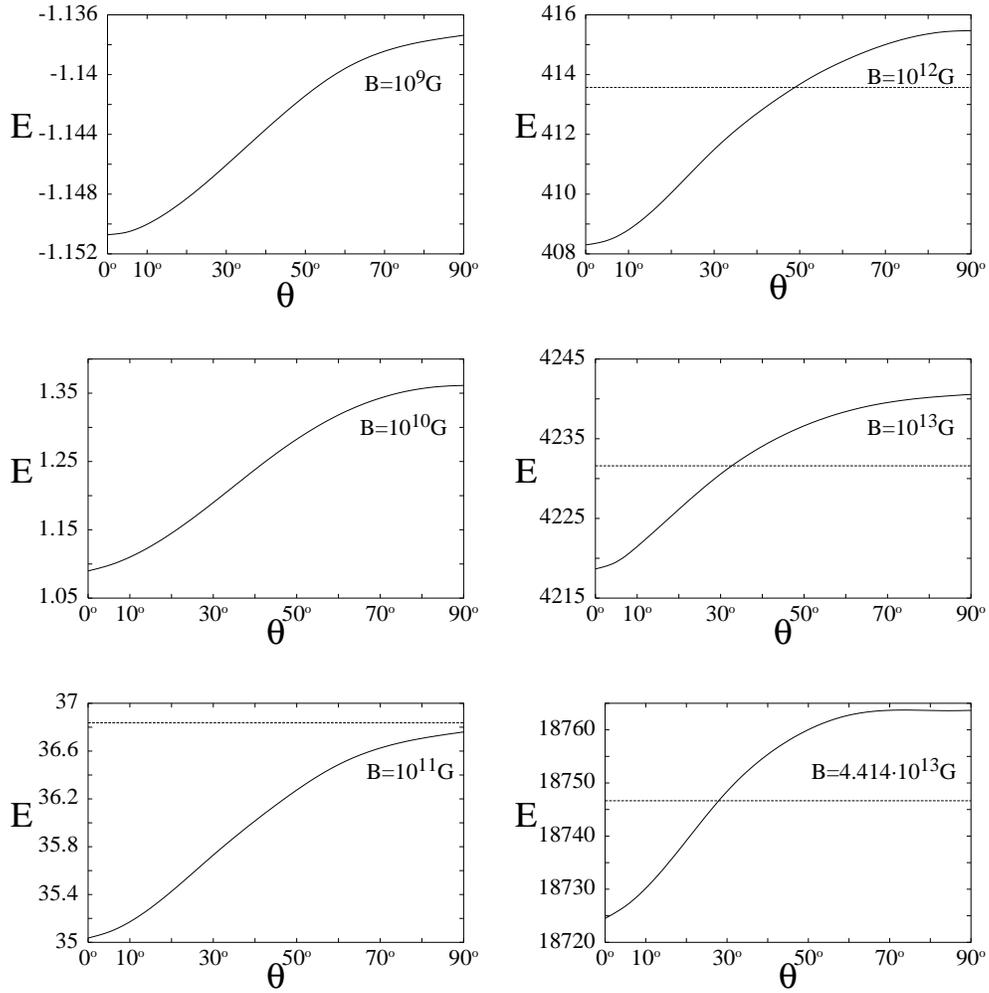,width=5.25in,angle=0}}
     \caption{Total energy of the $H_2^+$ ion as a function of the
       inclination angle. The horizontal lines refer to the $H$
       ground state energy taken from Lai et al. \cite{Salpeter:1992}. }
    \label{fig:3}
\end{center}
\end{figure}

\begin{figure}[htbp]
\begin{center}
  {\psfig{figure=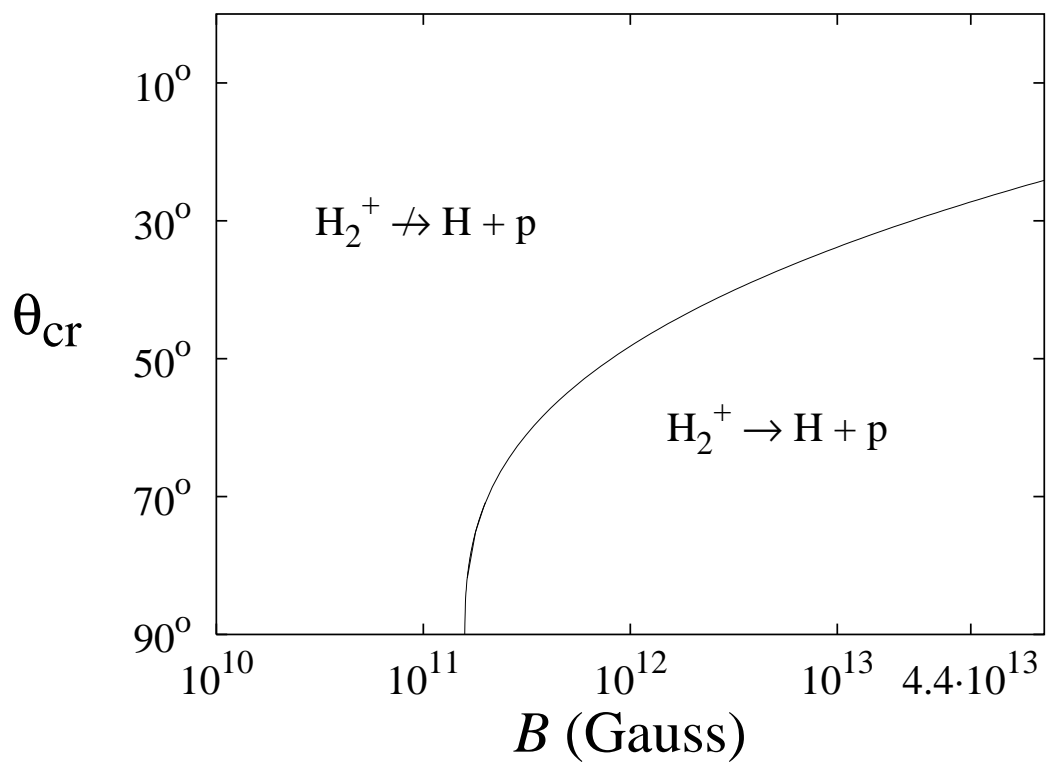,width=6.5in,angle=-90}} \caption{Critical
    angle for dissociation of  $H_2^+$.}
\end{center}
    \label{fig:4}
\end{figure}

\end{document}